\documentclass{pasj00}

\begin{document}
\SetRunningHead{Kurayama et al.}{Annual Parallax of MSXDC G034.43$+$00.24}

\title{Annual Parallax Measurements of an Infrared Dark Cloud MSXDC~G034.43$+$00.24 with VERA}

\author{Tomoharu \textsc{Kurayama},\altaffilmark{1}
	Akiharu \textsc{Nakagawa},\altaffilmark{1}
	Satoko \textsc{Sawada-Satoh},\altaffilmark{2}
	Katsuhisa \textsc{Sato},\altaffilmark{2}
	Mareki \textsc{Honma},\altaffilmark{2,3}
	Kazuyoshi \textsc{Sunada},\altaffilmark{2}
	Tomoya \textsc{Hirota},\altaffilmark{2,3} and
	Hiroshi \textsc{Imai}\altaffilmark{1}}
\altaffiltext{1}{Department of Physics and Astronomy, Graduate School of Science and Engineering, Kagoshima University, 1-21-35, Korimoto, Kagoshima, 890-0065}
\altaffiltext{2}{Mizusawa VLBI Observatory, National Astronomical Observatory of Japan, 2-12, Hoshigaoka-cho, Mizusawa, Oshu, Iwate, 023-0861}
\altaffiltext{3}{Department of Astronomical Sciences, Graduate University for Advanced Studies, 2-21-1, Osawa, Mitaka, Tokyo, 181-8588}

\email{kurayama@sci.kagoshima-u.ac.jp}



\maketitle

\begin{abstract}
	We have measured the annual parallax of the ${\rm H}_2{\rm O}$ maser source associated with an infrared dark cloud MSXDC G034.43$+$00.24 from the observations with VERA (VLBI Exploration of Radio Astrometry).  The parallax is $0.643 \pm 0.049$~mas, corresponding to the distance of $1.56^{+0.12}_{-0.11}$~kpc.  This value is less than the half of the previous kinematic distance of 3.7~kpc.  We revise the core mass estimates of MSXDC G034.43$+$00.24, based on virial masses, LTE masses and dust masses and show that the core masses decrease from the previous estimations of $\sim 1000 \MO$ to hundreds of $\MO$.  The spectral type derived from the luminosity also changes from O9.5 to B1 in the case of MM1.  This spectral type is still consistent with that of the massive star.  The radial velocity derived from the flat rotation model is smaller than the observed velocity, which corresponds to the peculiar motion of $\sim 40 {\rm ~km~s}^{-1}$ in the line-of-sight direction.
\end{abstract}

\section{Introduction}

Infrared dark clouds (hereafter, IRDCs) are discovered by the observations with ISO (Infrared Space Observatory) and MSX (Midcourse Space Experiment) \citep{bib:Perault+1996, bib:Carey+1998, bib:Hennebelle+2001}, and observed as dark silhouettes against the background radiation of our galaxy in the mid infrared wavelength \citep{bib:Perault+1996, bib:Egan+1998}.  They are massive ($\sim 10^2$--$10^4 \MO$), dense ($> 10^5~{\rm cm}^{-3}$), high column density ($\sim 10^{23}$--$10^{25}~{\rm cm}^{-2}$), and low temperature ($< 25$~K) \citep{bib:Egan+1998, bib:Carey+1998, bib:Carey+2000, bib:Tayssier+2002, bib:Simon+2006a, bib:Simon+2006b}.

Recently, IRDCs have been known as the sites where massive star formation is active \citep{bib:Rathborne+2006, bib:Rathborne+2007, bib:Pillai+2006, bib:Jackson+2008, bib:Chambers+2009}.  Various parameters, such as masses and luminosities, have been obtained from many observational studies from the mid-infrared to millimeter wavelengths (e.g., \cite{bib:Rathborne}; \cite{bib:Shepherd+2007}; \cite{bib:Sanhueza+2010}).  These parameters depend on distance, and in most cases, kinematic distances have been used to derive the parameters assuming the rotational model of our Galaxy.  However, recent annual parallax measurements with VLBI show that kinematic distances tend to be overestimated and sometimes more than the double of those from parallaxes (e.g. \cite{bib:Sato+2010}, \cite{bib:Motogi+}).  Therefore, it is important to measure distances toward the sources accurately to know the physical condition of the sources.

Annual parallaxes of galactic sources located at a few kpc from the Sun can be measured from the phase-referencing VLBI observations.  Phase-referencing VLBI is technique to observe two adjacent (less than several degrees in the case of 22~GHz-band observations) sources on the sky simultaneously or in the short time interval ($\lesssim 60$~s at 22 GHz band) \citep{bib:Asaki, bib:Thompson}.  By subtracting the difference of optical path lengths between the the target and reference source, it becomes possible to eliminate the short-term fluctuations of the earth's atmosphere.  In addition, the reference position is given by selecting an extragalactic source as the reference source.  These benefits enable the astrometry referred to the reference source and, with the long-time integration, the detections of faint sources which is not able to detect with the normal VLBI observations.

Phase-referencing VLBI observations with VERA (VLBI Exploration of Radio Astrometry) and the VLBA (Very Large Baseline Array) derive annual parallaxes of the kilo-parsec scale for the galactic maser sources (e.g., \cite{bib:baba} and references therein; \cite{bib:Sato+2010}; \cite{bib:Oh+2010}; \cite{bib:Rygl+2010}; \cite{bib:Kamohara+2010}) after the first successful measurements of annual parallaxes with ${\rm H}_2{\rm O}$ masers by \citet{bib:Kurayama2005}.  It is possible to measure the annual parallaxes when both the maser sources and the adjacent extragalactic reference sources are bright and compact sufficient to observe with VLBI.

 Some IRDCs emit masers such as the ${\rm H}_2{\rm O}$ masers and ${\rm CH}_3{\rm OH}$ masers \citep{bib:Pillai+2006, bib:VLA, bib:Chambers+2009}, so we can measure the annual parallaxes from the phase-referencing VLBI observations of these masers.  We used ${\rm H}_2{\rm O}$ masers this time because most parallaxes measured with VERA are based on the ${\rm H}_2{\rm O}$ maser observations.  \citet{bib:VLA} surveyed ${\rm H}_2{\rm O}$ masers for IRDCs and found that MSXDC G034.43$+$00.24 has bright ${\rm H}_2{\rm O}$ masers.  There is also an extragalactic source as a reference source for phase-referencing within $\timeform{2D}$ from MSXDC G034.43$+$00.24 \citep{bib:Fomalont+2003}.  Therefore, this source is one of the most suitable IRDCs to measure the parallaxes with VERA.

MSXDC G034.43$+$00.24 has a filamentary structure extended over $\sim \timeform{9'}$ in the declination direction, and four millimeter cores (MM1--MM4) on this filamentary structure \citep{bib:Faundez+2004, bib:Garay+2004, bib:Rathborne}.  An IRAS point source (IRAS 18507$+$0121) and an ultra compact H \emissiontype{II} region are associated with MM2 \citep{bib:Miralles+1994, bib:Molinari+1998}.  MM1, MM3 and MM4 have 4.5~\micron\ excesses, which suggest the existence of the ionized or shocked gas.  Point sources seen at 24~\micron\ and associated with all four millimeter cores suggest the existence of the warm dust \citep{bib:Rathborne, bib:Chambers+2009}.  \citet{bib:Garay+2004}, \citet{bib:Rathborne}, \citet{bib:Rathborne+2006}, \citet{bib:Shepherd+2007} and \citet{bib:Sanhueza+2010} calculated the mass of each millimeter core.  VLA observations have detected ${\rm H}_2{\rm O}$ masers in MM1, MM3 and MM4 \citep{bib:VLA}.  No VLBI observation has been conducted for MSXDC G034.43$+$00.24.  The available distances to MSXDC G034.43$+$00.24 is the kinematic distance only, whose value is 3.7~kpc, from \atom{C}{}{13}O $J = 1 \to 0$ observations \citep{bib:Simon+2006b} and CS(2--1) observations \citep{bib:Bronfman+1996, bib:Faundez+2004}.  Hence, it is very important to measure the annual parallax of this source.  Here, we report the results from VERA.

\section{Observations}

Observations were carried out with VERA four stations.  The target maser is the $6_{16} \rightarrow 5_{23}$ transition of ${\rm H}_2{\rm O}$ molecules, whose rest frequency is 22.23508 GHz.  We monitored at ten epochs from 2006 November to 2008 July with the typical observing duration of 9 hours.  The observation dates are listed in table \ref{table:obs-date}.
\begin{table*}
	\caption{Observation summary.}
	\label{table:obs-date}
	\begin{center}
		\begin{tabular}{rcrc} \hline
			epoch & date & day offset\footnotemark[$*$]
				& synthesized beam\footnotemark[$\dagger$] \\ \hline
			1st  & 2006 Nov 16 & $-297$ & $1.29 \times 0.86$ mas at $\timeform{-53D}$ \\
			2nd  & 2007 Jan 04 & $-248$ & $1.40 \times 0.79$ mas at $\timeform{-50D}$ \\
			3rd  & 2007 Feb 10 & $-211$ & $1.39 \times 0.82$ mas at $\timeform{-52D}$ \\
			4th  & 2007 Mar 25 & $-168$ & $1.36 \times 0.81$ mas at $\timeform{-52D}$ \\
			5th  & 2007 May 04 & $-128$ & $1.28 \times 0.74$ mas at $\timeform{-53D}$ \\
			6th  & 2007 Aug 06 & $- 34$ & $1.32 \times 0.81$ mas at $\timeform{-44D}$ \\
			7th  & 2007 Oct 02 &    23  & $1.33 \times 0.80$ mas at $\timeform{-45D}$ \\
			8th  & 2008 Jan 04 &   117  & $1.43 \times 0.79$ mas at $\timeform{-49D}$ \\
			9th  & 2008 Apr 12 &   216  & $1.35 \times 0.82$ mas at $\timeform{-48D}$ \\
			10th & 2008 Jul 02 &   297  & $1.27 \times 0.81$ mas at $\timeform{-39D}$ \\ \hline
			\multicolumn{4}{@{}l@{}}{\hbox to 0pt{\parbox{180mm}{\footnotesize
				\footnotemark[$*$] Offset from the center date (2007 September 9) of the term.  \par\noindent
				\footnotemark[$\dagger$] The lengths of major and minor axes and the position angle of the major axis.
			}\hss}}
		\end{tabular}
	\end{center}
\end{table*}
Before this astrometric monitoring, we conducted maser survey observations of four millimeter cores on 2006 October 21.

We conducted phase-referencing VLBI observations of the target source MSXDC G034.43$+$00.24 and the reference source GPSR5 35.946$+$0.379 ($=$ VCS2 J1855$+$0251, hereafter we call this source as the reference source), which is separated by $\timeform{1.6D}$ from the target source.  Using the dual beam system of VERA telescopes \citep{bib:2BKawaguchi, bib:2bcal}, we observed these two sources simultaneously.  For the fringe finding and bandpass calibration, we also observed the calibrator source QSO J1800$+$3848 ($\alpha_{{\rm J2000}} = \timeform{18h00m24.765362s}$, $\delta_{{\rm J2000}} = \timeform{+38D48'30.69754"}$) with the both beams every 80 minutes.

Observed signals are filtered with the digital filter bank \citep{bib:DFU} and outputted in the form of sixteen IFs.  Each IF contains 16-MHz bandwidth signal.  One IF contains the signal of the target (maser) source.  The other fifteen IFs contain the signals of the reference (continuum) source.  The recording rate is 1024 Mbps ($32 {\rm ~ Msps} \times 2 {\rm ~bit~sample}^{-1} \times 16 {\rm ~ IF}$).  The correlation is performed with Mitaka FX correlator.  The correlator integrates the output data for 1~s in the time domain.  In the frequency domain, the output data are 8~MHz bandwidth with the 512 frequency points for one maser IF and 16~MHz bandwidth with the 64 frequency points for fourteen continuum IFs.  One continuum IF is abandoned in the data reduction process.  The velocity resolution at the correlator output is $0.21~{\rm km~s}^{-1}$ for the maser IF and $3.4~{\rm km~s}^{-1}$ for the continuum IFs.

The measurements of the annual parallax of MSXDC G034.43$+$00.24 require a special care because this source has a low-declination of $\sim + \timeform{1D}$.  This causes a decrease of the positional accuracy in the declination direction.  In VLBI observations, we vary the measurement points on the $uv$ plane, which corresponds to the aperture of the normal single-dish telescopes, by changing the positional relationship between the stations and the sources using the earth's spin.  The loci of measured points on the $uv$ plane draw parts of the ellipses for high-declination sources, but they draw almost straight lines in the right-ascension direction for low-declination sources.  Thus the positional accuracy in the declination direction decreases especially in the case of the small number of stations, that is, small numbers of baselines and observing points in the declination direction.  In our parallax measurement, we used only the right-ascension components in order to avoid this effect.  The reason of large uncertainty in the declination direction is also described in \S\ref{section:airmass}.

\section{Data Reduction}
\label{section:rough-analysis}

Data reduction was done with NRAO (National Radio Astronomy Observatory) AIPS (Astronomical Image Processing System) software and DIFMAP.  The functions neither AIPS nor DIFMAP provide were added with external programs.
The steps of our data reduction are as follows :
\begin{enumerate}
\item
	Amplitude calibration for all sources
\item
	Bandpass calibration for all sources using calibrator source
\item
	Modification of tracking model in Mitaka FX correlator
	\label{list:recalc}
\item
	Time and frequency integration
\item
	Calibration of clock parameters (Global fringe search for calibrator source)
\item
	Global fringe search for reference source
\item
	Imaging of reference source with self-calibration
\item
	Phase referencing (Subtracting the phases of reference source from those of target source)
\item
	Dual-beam calibration with `horn-on-dish' method \citep{bib:2bcal}
\item
	Calibration of Doppler effect
\item
	Imaging of target source with CLEAN
	\label{list:imaging}
\item
	Measuring the position and flux by fitting elliptical Gaussians
	\label{list:positioning}
\end{enumerate}
The details of the data reduction procedure for an observing epoch are shown in Appendix \ref{section:analysis}.

`Modification of tracking model in Mitaka FX correlator' (step \ref{list:recalc}) is needed because the tracking model in Mitaka FX correlator does not have sufficient accuracy for astrometric measurements.  We changed it to the newly calculated model with CALC3/MSOLV \citep{bib:JikeFXCALC, bib:ManabeFXCALC}.  In this recalculation, we used the equatorial coordinates for the target source MSXDC G034.43$+$00.24 as $\alpha_{{\rm J2000}} = \timeform{18h53m19.00000s}$, $\delta_{{\rm J2000}} = \timeform{+1D24'08.0000"}$ and for the reference source GPSR5 35.946$+$0.379 $=$ VCS2 J1855$+$0251 as $\alpha_{{\rm J2000}} = \timeform{18h55m35.43649s}$, $\delta_{{\rm J2000}} = \timeform{+2D51'19.5623"}$ \citep{bib:Fomalont+2003}.  The equatorial coordinate for the target source corresponds to that of the millimeter core MM1 \citep{bib:Rathborne}.  The new recalculated model contains the estimation of the wet component of the earth's atmosphere from GPS data at VERA stations \citep{bib:secz}.

\begin{figure}
	\begin{center}
		\FigureFile(80mm, 80mm){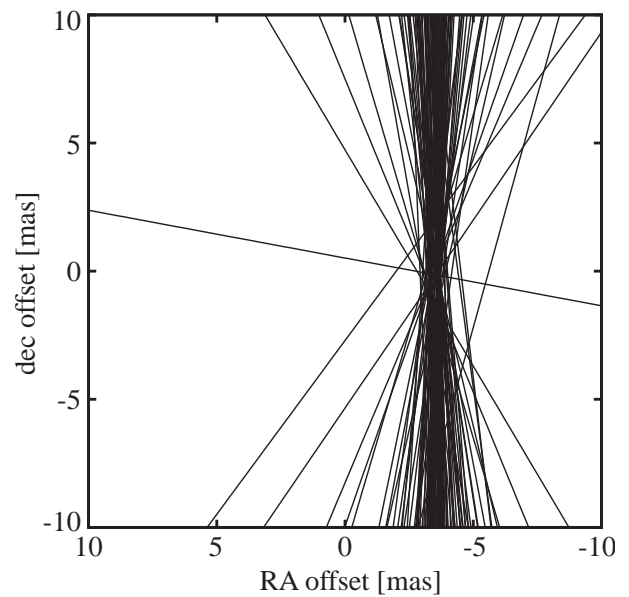}
	\end{center}
	\caption{Fringe rate map of the ${\rm H}_2{\rm O}$ masers at the millimeter core MM1 made from the results of the global fringe search with AIPS.  The map origin is located at the position of the target source modified in the recalculation of the tracking model whose equatorial coordinate is $\alpha_{{\rm J2000}} = \timeform{18h53m19.00000s}$, $\delta_{{\rm J2000}} = \timeform{+1D24'08.0000"}$.  Each line shows the possible position of the masers calculated from one rate value.  Maser clusters are placed at the crossing point of lines.  This map is obtained from the data on of 2008 January 4 at the radial velocity of $v_{\rm LSR} = 56.5~{\rm km~s}^{-1}$.}
	\label{fig:rate-map}
\end{figure}

Before `imaging of target source with CLEAN' (step \ref{list:imaging}), we need to know the position where we make images.  Only the ${\rm H}_2{\rm O}$ masers associated with the millimeter core MM1 was imaged (see \S\ref{section:survey}).  Producing the fringe-rate map with our own program from the results of the global fringe search for the flux-peak frequency chennel, we find a maser cluster as shown in figure \ref{fig:rate-map}.
We made the images for the eight features in the maser cluster at which we found the masers for the data on 2008 January 4.  We adopted maser features only when their radiation is confirmed over three or more frequency channels.

\begin{figure}
	\begin{center}
		\FigureFile(80mm, 90mm){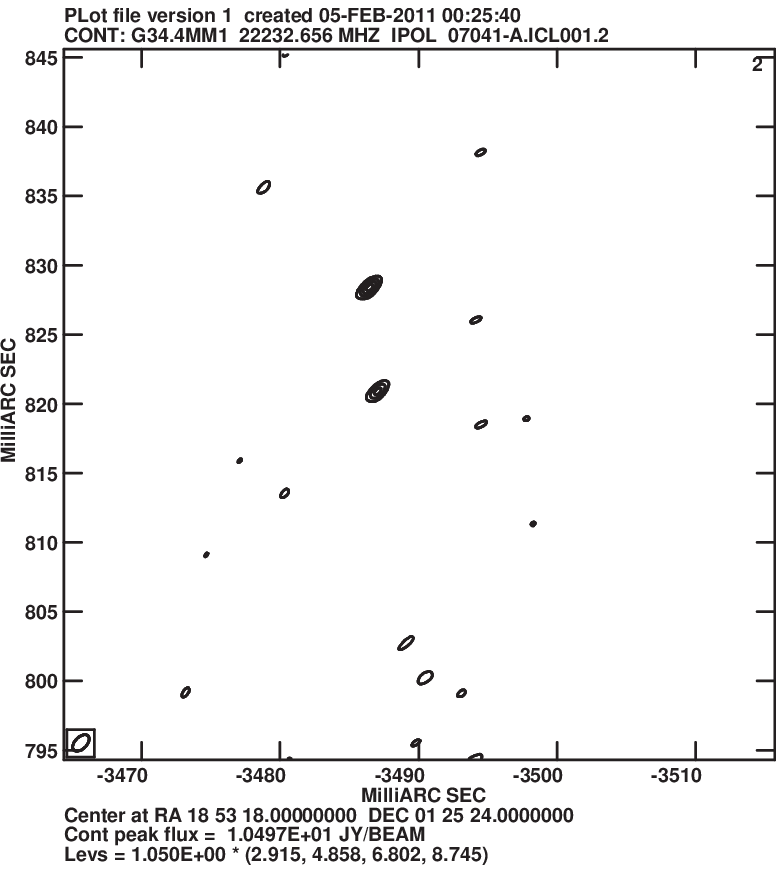}
	\end{center}
	\caption{An example of the phase-referenced images of detected maser features.  The phase-referenced image of maser feature No.\ 2 at 3rd epoch on 2007 Feburuary 10.  Contours are $3\sigma$, $5\sigma$, $7\sigma$ and $9\sigma$ brightness level, where $\sigma$ is the rms noise level of $0.99 {\rm ~Jy~beam}^{-1}$.  Maser features are compact sufficient to measure their positions.}
	\label{fig:maser-map}
\end{figure}

\begin{figure}
	\begin{center}
		\FigureFile(80mm, 120mm){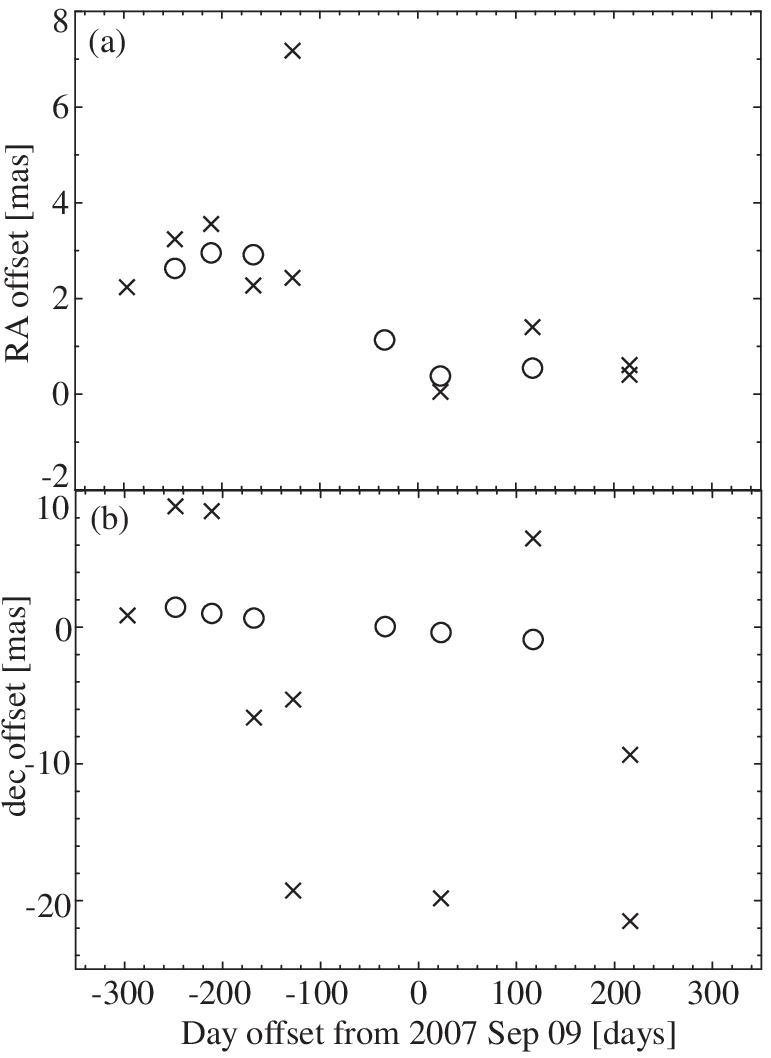}
	\end{center}
	\caption{Maser-feature selection for the feature No.\ 2 when multiple maser features are detected in an image at an epoch.  (a) Plot of right ascensions versus time.  (b) Plot of declinations versus time.  In both panels, open circles ($\bigcirc$) show the maser features which are used for the parallax fitting, and crosses ($\times$) show the maser features which are not used for the parallax fitting.  From (a), it is difficult to know which feature should be used, but we can easily select from (b) although declination data have large position errors shown in \S\ref{section:airmass}.}
	\label{fig:selection}
\end{figure}

In `measuring the position and flux by fitting elliptical Gaussians' (step \ref{list:positioning}), we fit the elliptical Gaussians for the five regions where we detected maser emission during more than two continuous epochs (see, table \ref{table:detect}).  We adopted the peak positions as the positions of maser features.  In this paper, maser features denotes maser emissions detected at multiple frequency points at the same position on the sky.  Positions are measured at the frequency channels at which the flux is largest among the detected frequency channels.  Detected maser features are compact sufficient for the position measurements, shown in Figure \ref{fig:maser-map}.
When multiple maser features are detected in an image at an epoch, we used the feature which has closest declination value to the other epochs.  Figure \ref{fig:selection} shows an example of the selection of maser features in this process.
From this figure, we can find that declinations are better for this selection than right ascensions although declinations have large position errors shown in \S\ref{section:airmass}.

\section{Results}

\subsection{Survey Results of Masers}
\label{section:survey}

\begin{figure*}
	\begin{center}
		\FigureFile(150mm, 150mm){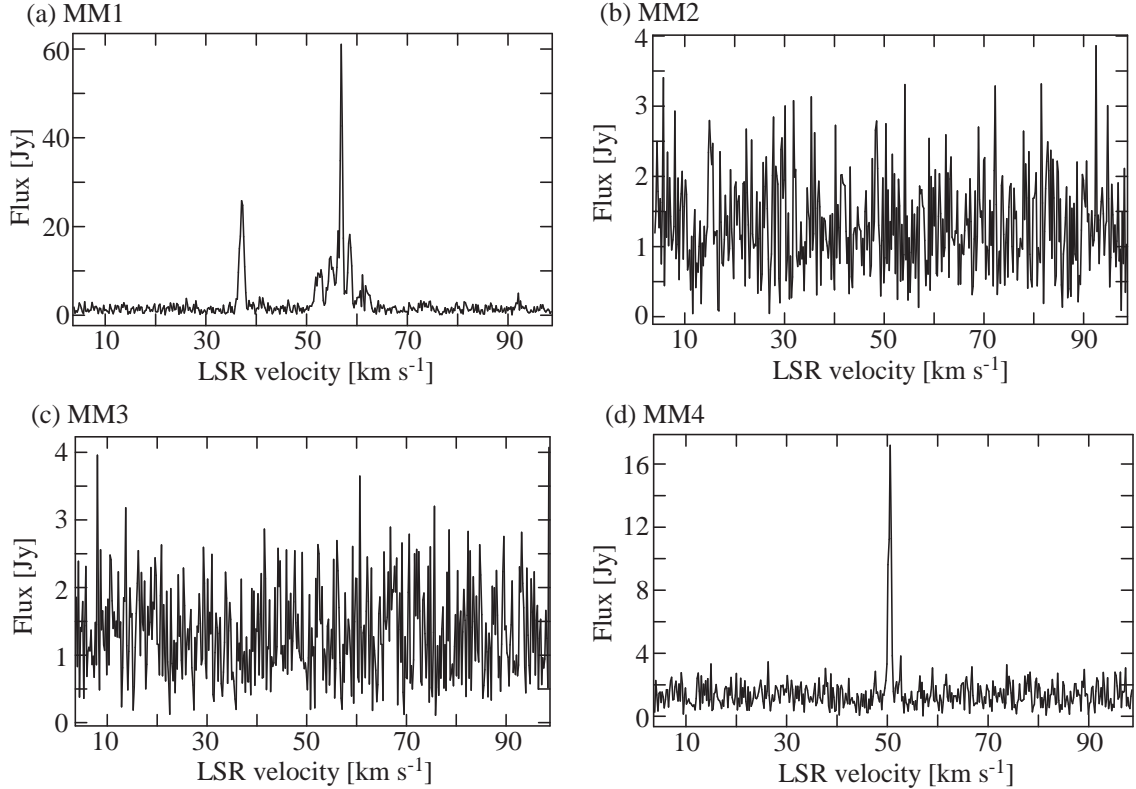}
	\end{center}
	\caption{Cross-power spectra of the ${\rm H}_2{\rm O}$ masers at the millimeter-core positions: (a) MM1, (b) MM2, (c) MM3, and (d) MM4.  They are obtained from the data on 2006 October 21 at Mizusawa--Iriki baseline.  The integration time of data is 5 minutes.  We integrated the data after the phase-referencing, which gets rid of the short-time fluctuation of the earth's atmosphere and enables the long-term integration.}
	\label{fig:cross-spectr}
\end{figure*}

Figure \ref{fig:cross-spectr} shows the cross-power spectra obtained from the survey observations on 2006 October 21 at the four millimeter-core positions.
We find that the ${\rm H}_2{\rm O}$ masers are detected in MM1 and MM4, but are not detected in MM2 and MM3.  Comparing with the VLA observations in 2006 \citep{bib:VLA}, the detection is not consistent only for MM3.  This may be caused by the time variation of masers or the extended emission of the source structure, that is, masers in MM3 could be resolved out with the long ($\sim 2000$~km) baselines of VERA.

From this result, we observed the millimeter-core MM1 for the annual parallax measurements from the observation on 2006 November 16.

\subsection{Derivation of Parallax with Least Square Fitting}
\label{section:fitting}

\begin{figure}
	\begin{center}
		\FigureFile(80mm, 60mm){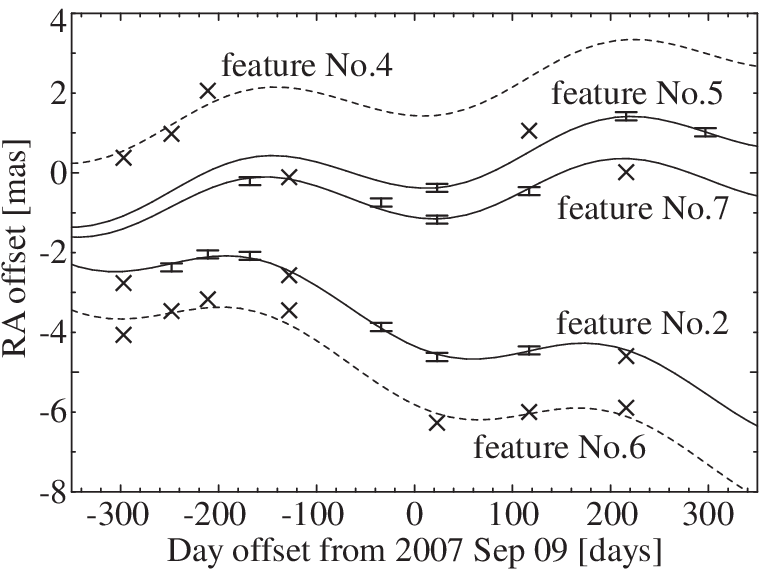}
	\end{center}
	\caption{Observed data points in the right ascension direction and the least square fitting for the annual parallax measurements.  Table \ref{table:obs-date} shows the correspondence between the values of day offsets from 2007 September 9 (horizontal axis) and the observation dates.  The offset in the vertical axis for each feature is included just for the display.  The observed points are the center of the error bars.  Crosses ($\times$) show the data points which are not used for the parallax fitting.  The amplitudes of uncertainty are 0.101~mas, which is calculated from the assumption of $({\rm reduced} ~ \chi^2) = \chi^2 / \nu = 1$.  Feature numbers are same as those in table \ref{table:detect}.  Solid curves are the results of the least-square fitting as shown in table \ref{table:fit-result}.  Dashed curves show the movements of features which are derived from the least square fitting for calculating linear motions and initial positions only, assuming the parallax of $\varpi = 0.643$~mas derived from the fitting.}
	\label{fig:fitting}
\end{figure}

\begin{figure}
	\begin{center}
		\FigureFile(80mm, 60mm){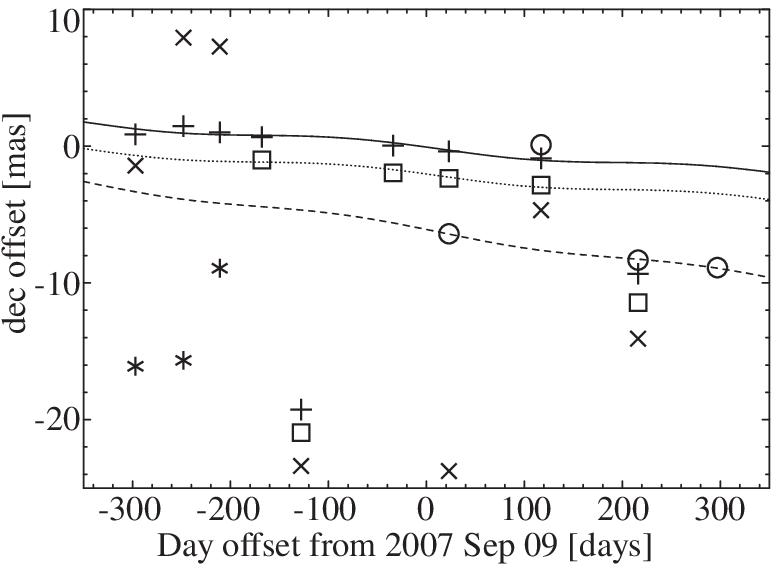}
	\end{center}
	\caption{Plot of the observed data points in the declination direction against time.  Table \ref{table:obs-date} shows the correspondence between the day offsets from 2007 September 9 (horizontal axis) and the observation dates.  The offset in the vertical axis is included for each feature for the display.  Crosses ($+$) show the positions of feature number 2, asterisks ($*$) of feature number 4, circles ($\bigcirc$) of feature number 5, Crosses ($\times$) of feature number 6 and squares ($\square$) of feature number 7.  Feature numbers are same those in table \ref{table:detect}.  Solid, dotted and dashed curves show the movements of feature number 2, 7 and 5 respectively, which are derived from the least square fitting for calculating linear motions and initial positions only, assuming the parallax derived from right-ascention data $\varpi = 0.643$~mas.}
	\label{fig:dec}
\end{figure}

\begin{figure}
	\begin{center}
		\FigureFile(80mm, 80mm){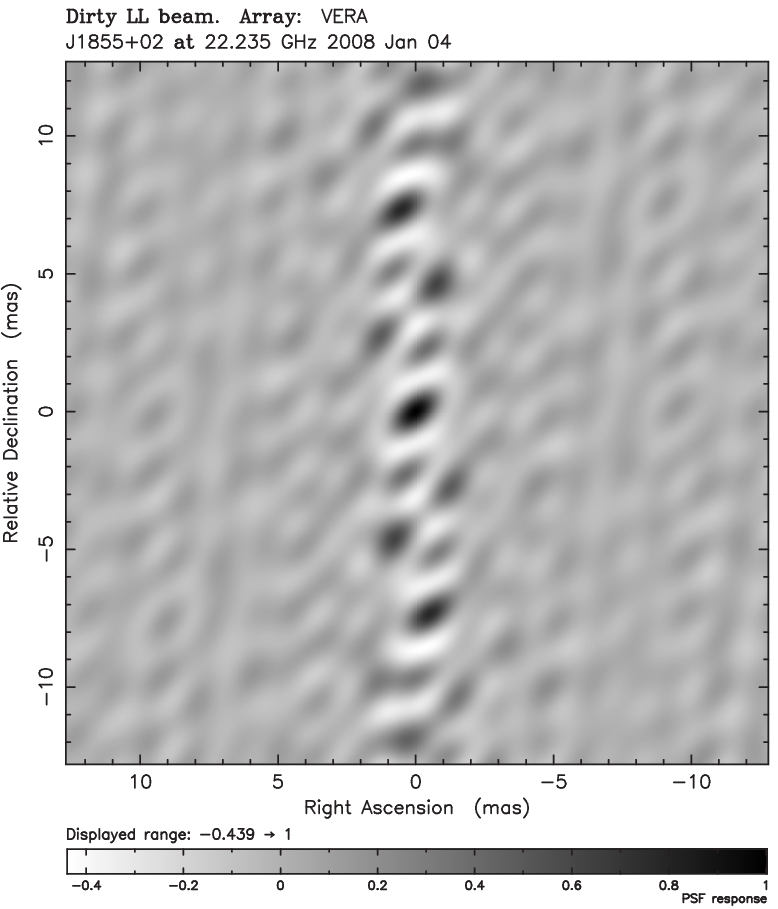}
	\end{center}
	\caption{The synthesized beam of VERA four stations in the observation on 2008 January 4.  Strong sidelobes are found in the declination direction.}
	\label{fig:synth-beam}
\end{figure}

Figure \ref{fig:fitting} and \ref{fig:dec} show the movements of maser features measured with respect to the reference source.
The movements look like the combination of the parallax motion and the linear motion in the right-ascension direction, but have large scatter over 20~mas in the declination direction.  This is because the declination of the target source ($\sim + \timeform{1D}$) is small.  As shown in figure \ref{fig:synth-beam}, there are large sidelobes of the synthesized beam in the declination direction.  The stable and uncalibrated component of the earth's atmosphere affect the positions in the declination direction.
The details of this effect is discussed in \S\ref{section:airmass}.  We used only the data of the right-ascension direction this time.

We carried out the least square fitting only with the right-ascension data to derive the annual parallax.  The model is the combination of the movement by the annual parallax and the linear motion.  The movement except by the annual parallax, such as the galactic rotation, the peculiar motion of the Sun, are contained in the linear motion.  This is expressed as the following equation:
\begin{eqnarray}
	\alpha^{(i)}(t) \cos\delta
	& = & \varpi (- \sin\alpha \cos\solar + \cos\varepsilon \cos\alpha \sin\solar) \nonumber \\
	& & ~~~ + (\mu_{\alpha}^{(i)} \cos\delta) t + \alpha^{(i)}_0 \cos\delta,
\end{eqnarray}
where $\alpha^{(i)}(t)$ is the observed right ascension of the $i$'th maser feature at the time $t$, $\alpha$ and $\delta$ are the right ascension and declination of the target source respectively, $\varpi$ is the annual parallax, $\solar$ is the ecliptic longitude of the Sun and $\varepsilon$ is the obliquity of ecliptic.  $\mu_{\alpha}^{(i)} \cos\delta$ is the right-ascension component of the linear motion of the $i$'th maser feature, $t$ is the day offset from the center of the observation term used for the parallax measurements, shown in table \ref{table:obs-date} and $\alpha^{(i)}_0$ is the right ascension of $i$'th maser feature at the time $t = 0$ \citep{bib:Green1985}.

Table \ref{table:detect} summarizes the detection of masers at each feature and at each epoch.
\begin{table*}
	\caption{Detection at each feature and epoch.}
	\label{table:detect}
	\begin{center}
		\begin{tabular}{cccccccccccccc} \hline
			Feature & $v_{{\rm LSR}}$ & RA\footnotemark[$*$] & dec\footnotemark[$*$] & \multicolumn{10}{c}{Integrated flux at each epoch [Jy]\footnotemark[$\dagger$]} \\
			Number & [${\rm km~s}^{-1}$] & [arcsec] & [arcsec] & 1st & 2nd & 3rd & 4th & 5th & 6th & 7th & 8th & 9th & 10th \\ \hline
			1 & 88.7 & $-3.632$ & $+0.088$ & \ldots & (1.4) & (1.1) & \ldots & \ldots & \ldots & (2.6) & (3.6) & \ldots & \ldots \\
			2 & 61.7 & $-3.489$ & $+0.819$ & (3.4) & 5.9 & 8.8 & 3.8 & (5.7) & 3.1 & 2.5 & 9.4 & (2.8) & \ldots \\
			3 & 59.4 & ($-3.367$) & ($-0.036$) & \ldots & (2.1) & \ldots & \ldots & \ldots & \ldots & (5.2) & \ldots & (2.2) & \ldots \\
			4 & 57.9 & $-3.425$ & $-0.061$ & (2.4) & (4.4) & (3.9) & \ldots & \ldots & \ldots & (0.8) & (2.0) & \ldots & \ldots \\
			5 & 57.3 & $-3.303$ & $-0.035$ & (1.7) & (3.0) & \ldots & \ldots & \ldots & \ldots & 1.0 & (5.3) & 22.4 & 11.5 \\
			6 & 56.5 & $-3.672$ & $+0.107$ & (3.2) & (6.6) & (5.8) & (4.5) & \ldots & \ldots & (2.6) & (27.3) & (54.5) & \ldots \\
			7 & 53.9 & $-3.425$ & $+0.657$ & \ldots & \ldots & \ldots & 2.2 & (8.7) & 3.8 & 8.7 & 19.2 & (5.8) & \ldots \\
			8 & 43.4 & $-3.392$ & $-0.048$ & \ldots & \ldots & \ldots & \ldots & \ldots & \ldots & \ldots & (5.1) & (0.5) & \ldots \\ \hline
			\multicolumn{14}{@{}l@{}}{\hbox to 0pt{\parbox{180mm}{\footnotesize
				\footnotemark[$*$] Offset values from the right ascension and declination used in the modification of the tracking model : $\alpha_{{\rm J2000}} = \timeform{18h53m19.0s}$, $\delta_{{\rm J2000}} = \timeform{+1D24'08"}$.  Values without parentheses are the offsets at 8th epoch on 2008 January 4.  Values with parentheses are the averages of the detected epochs. \par\noindent
				
				\footnotemark[$\dagger$] Numbers without parentheses denote that the data of the features and epochs are used for the annual parallax measurements.  Numbers with parentheses denote that the data are not used.
			}\hss}}
		\end{tabular}
	\end{center}
\end{table*}
Maser features detected during more than two continuous epochs are used for the parallax measurements.  The data points which are not on the declination trend are omitted from the fitting.  This is because we can not solve parameters when we can not use data of more than two epochs.  When we include one new maser feature, we need two extra fitting parameters: the initial position and the linear motion.  We fitted for three maser features, so seven parameters are fitted : one common $\varpi$, five $\mu_{\alpha}^{(i)}$, five $\alpha^{(i)}_0 \cos\delta$.  The total number of data points is twenty-nine, as shown in table \ref{table:detect}.

Figure \ref{fig:fitting} and table \ref{table:fit-result} show the results of fitting.
\begin{table}
	\caption{Results of the fitting of the annual parallax and the linear motions}
	\label{table:fit-result}
	\begin{center}
		\begin{tabular}{crr} \hline
			$\varpi$\footnotemark[$*$] & \multicolumn{2}{l}{$0.643 \pm 0.049$~mas} \\ \hline \hline
			Feature\footnotemark[$\dagger$] & \multicolumn{1}{c}{$\mu_{\alpha}^{(i)} \cos\delta$\footnotemark[$\ddagger$]} & \multicolumn{1}{c}{$\alpha_0^{(i)} \cos\delta$\footnotemark[\S]} \\
			Number & \multicolumn{1}{c}{[${\rm mas~yr}^{-1}$]} & \multicolumn{1}{c}{[mas]} \\ \hline
			2 & $-2.19 \pm 0.12$ & $-3488.77 \pm 0.05$ \\
			5 & $ 0.98 \pm 0.21$ & $-3303.80 \pm 0.12$ \\
			7 & $ 0.46 \pm 0.19$ & $-3425.55 \pm 0.05$ \\ \hline
			\multicolumn{3}{@{}l@{}}{\hbox to 0pt{\parbox{85mm}{\footnotesize
				\footnotemark[$*$]
					Value of the annual parallax. \par\noindent
				\footnotemark[$\dagger$]
					Corresponding to the feature numbers in table \ref{table:detect}. \par\noindent
				\footnotemark[$\ddagger$]
					Right-ascension components of the linear motions. \par\noindent
				\footnotemark[\S]
					Offsets in the right-ascension direction at $t = 0$ on 2007 September 9 from the position used in the modification of tracking model $\alpha_{\rm J2000} = \timeform{18h53m19.0s}$, $\delta_{\rm J2000} = \timeform{+1D24'08"}$.
			}\hss}}
		\end{tabular}
	\end{center}
\end{table}
The derived annual parallax is $0.643 \pm 0.049$~mas, corresponding to the distance of $1.56^{+0.12}_{-0.11}$~kpc.  This distance is smaller than the half of the previous kinematic distance of 3.7~kpc.

\subsection{Uncertainty of Each Epoch}

Uncertainty at one epoch is calculated by assuming that the uncertainty values of all epochs are same and that $({\rm reduced} ~ \chi^2) = \chi^2 / \nu = 1$, where $\chi^2$ is the squared sum of the residuals of fitting over the square of the uncertainty of one epoch, $\nu$ is the degree of freedom in the fitting, the total number of data points minus the total number of parameters.  In our fitting, $\nu = 29 - 11 = 18$.  The uncertainty at one epoch is 0.101~mas.  The uncertainty of fitted parameters is calculated from this value with the normal least square fitting method.  Most of the uncertainty is the stable component of the earth's atmosphere of about 0.2~mas, as discussed in \S\ref{section:airmass}.

\begin{figure}
	\begin{center}
		\FigureFile(80mm, 100mm){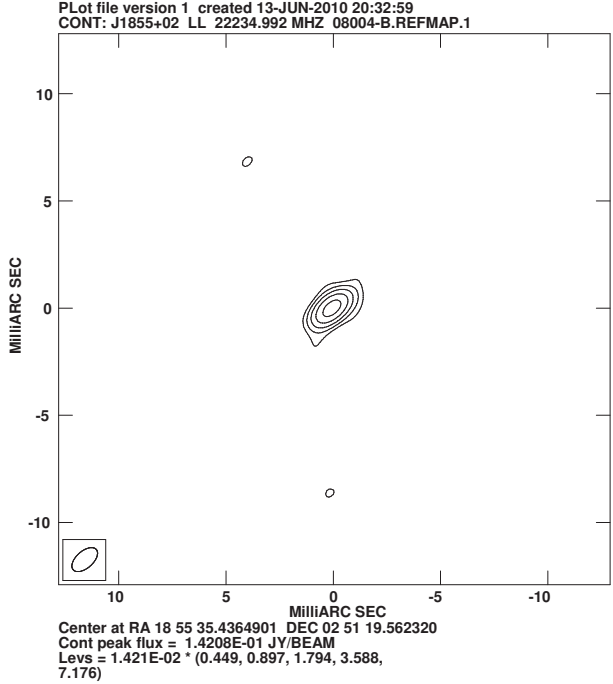}
	\end{center}
	\caption{Imaging result of the reference source GPSR5 35.946$+$0.379 $=$ VCS2 J1855$+$0251 with VERA four stations at 22 GHz band.  The observation date is 2008 January 4.  Contours are $3 \sigma$, $6 \sigma$, $12 \sigma$, $24 \sigma$ and $48 \sigma$ brightness level, where $\sigma$ is the rms noise level of $2.1~{\rm mJy~beam}^{-1}$.}
	\label{fig:ref-map}
\end{figure}

The uncertainty from the structure of the reference source is small.  Figure \ref{fig:ref-map} shows the imaging result of the reference source.
This source structure has little change over the all observation epochs used for the parallax measurements.  We used same stations in all observing epochs, so the measurement points on the $uv$ plane are basically same.

In order to evaluate the uncertainty of the parallax measurements from the structure of the reference source, we calculated the worst value of the uncertainty as follows.  The delay time, optical path length divided by the speed of light, caused by the source structure $\tau(u, v, \omega)$ is given by
\begin{equation}
	\tau(u, v, \omega) = \frac{d}{d\omega} \left\{ \arg \left[ \iint B(x, y) e^{- 2\pi i (ux + vy)} \, dx dy \right] \right\},
\end{equation}
where $B(x, y)$ is the brightness distribution of the reference source, $x$ and $y$ are the coordinates of the right-ascension and declination direction respectively, $\omega = 2\pi\nu$ and $\nu$ is the observing frequency \citep{bib:structure1, bib:structure2}.  For the simple calculations, we set $B(x, y)$ as the sum of the point sources obtained from CLEAN, $B(x, y) = \sum_i B_{0i} \delta(x_i, y_i)$, and approximate $d/d\omega \sim 1 / \omega = 1 / (2\pi\nu)$ and $2\pi\nu\tau(u, v, \omega) \sim 2\pi \sqrt{u^2 + v^2} \Delta x$, where $\Delta x$ is the positional error.  Then, so we approximate \mbox{$2\pi\nu\tau(u, v, \omega) \sim 2\pi\sqrt{u^2 + v^2} \Delta x$} where \mbox{$\Delta x$} is the positional error.  Then
\begin{equation}
	\Delta x \sim \frac{1}{2\pi\sqrt{u^2 + v^2}} \arg \left( \sum_i B_{0i} e^{- 2\pi i (ux_i + vy_i)} \right).
\end{equation}
We calculated this value for the data on 2008 January 4 and the maximum values of $u$ and $v$ of $u_{\max} = 1.4 \times 10^8$ and $v_{\max} = 1.2 \times 10^8$ in the unit of the wavelength.  The result is $\Delta x = 0.045~{\rm mas}$.  The uncertainty from the structure of the reference source has a small fraction in the each epoch's uncertainty of 0.101~mas.

\subsubsection{Uncertainty from Stable Component of the Earth's Atmosphere}
\label{section:airmass}

In the phase-referencing VLBI observations, the effect of the earth's atmosphere is removed by taking the differences of the optical path lengths between the target and reference sources.  With this process, we can remove the short-term fluctuations of the earth's atmosphere, which is the largest atmospheric effect in VLBI observations.  However, the elevation angle of the target and reference sources are not completely same, so it is impossible to remove the effect of airmass, that is, the stable component of the earth's atmosphere \citep{bib:secz}.  Assuming the plane-parallel approximation of the earth's atmosphere, the effect of airmass is given by in the unit of the optical path length
\begin{equation}
	l_{ij} = L_j \sec z_{ij},
\end{equation}
when source $i$ is observed with the station $j$, where $L_j$ is the airmass effect in the zenith direction in the unit of the optical path length, and $z_{ij}$ is the zenith angle of the observed source.  In phase-referencing observations, we subtract the optical path lengths between the target and reference sources.  After the phase-referencing, the following airmass effect remains:
\begin{equation}
	l_{1j} - l_{2j} = L_j (\sec z_{1j} - \sec z_{2j}).
\end{equation}

We evaluated quantitatively how much this effect contributed to the uncertainty of our astrometric observations in the following way: (1) Giving the optical-path-length offset in the zenith direction by the earth's atmosphere, $L$.  We added this offset to Ishigakijima station only for the simplicity because this station is most humid and hence, $L$ could have large uncertainty.  We changed $L$ at 1~cm step in the range of $- 30 {\rm ~ cm} \leq L \leq 30 {\rm ~ cm}$.  (2) Adding the phase correction of
\begin{equation}
	\frac{2 \pi}{\lambda} L [ \sec z_t(t) - \sec z_r(t) ]
\end{equation}
obtained from the plane-parallel approximation to all visibilities of the target source at the baselines including Ishigakijima station, where $\lambda$ is the observation wavelength, $z_t(t)$ and $z_r(t)$ are the zenith angle of the target and reference source at the time $t$, respectively.  We used the relationship between the phase $\phi$ and the optical path length $l$ of $\phi = (2 \pi / \lambda) l$.  (3) Imaging the target source from the phase-corrected visibilities without CLEAN, and deriving the flux and position of the maser feature by picking up the pixel of the flux peak.

\begin{figure}
	\begin{center}
		\FigureFile(80mm, 180mm){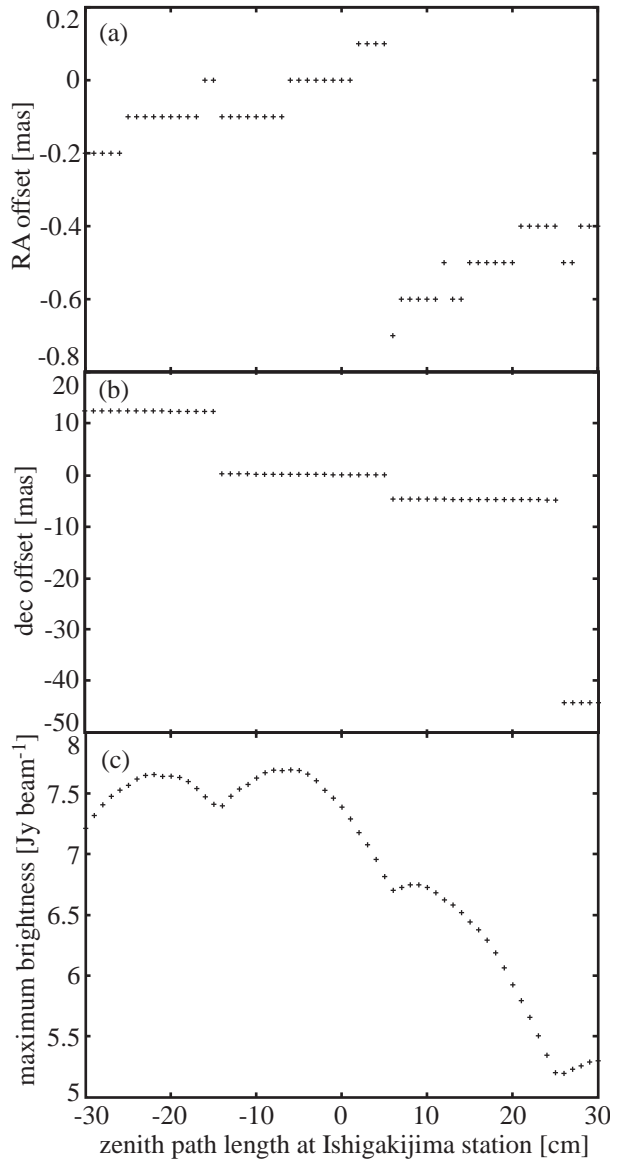}
	\end{center}
	\caption{(a) Right-ascension offset, (b) declination offset, (c) brightness of the peak-flux pixel against the optical path length of the zenith direction by the earth's atmosphere.  In (a) and (b), the right ascension and declination not added the atmospheric effect are set to zero.  We evaluated for the maser feature in the feature number 5 observed on 2008 January 4.  Since the pixel size is 0.1~mas and since the resolution of right-ascension and declination is the pixel size, the quantized variation in (a) is caused by the pixel size.  The  nature of quantized variation in (b) may be the effect of the sidelobe pattern (see figure \ref{fig:synth-beam}).}
	\label{fig:airmass}
\end{figure}

Figure \ref{fig:airmass} shows the result of above evaluation for the maser feature in the feature number 5 and for the observation on 2008 January 4.
As discussed by \citet{bib:secz}, the true atmospheric path length should maximize the peak brightness.  There are not so large offset in the right-ascension and declination direction at the atmospheric path length of about $- 8$~cm, but the offset becomes large, more than 10~mas in the declination direction at the atmospheric path length of around $- 22$~cm.  Such a large $L$ would not be expected if GPS calibration is included \citep{bib:secz} in which case it is normally a few centimeters.  Therefore, the stable component of the earth's atmosphere might be one of the largest causes of the large position offsets in the declination direction seen in figure \ref{fig:fitting}, but we can not deny other causes, for example the structure effect of the target maser source, large offsets ($\sim \timeform{4"}$) from the right ascension of the tracking model.  This atmospheric effect is the most part of the uncertainty in right ascension direction at each epoch of 0.101~mas, because the offset value in the right-ascension direction in the range of $- 25 ~ {\rm cm} < L < 0 ~ {\rm cm}$ is about $\pm 0.1$~mas.

\subsection{Other Approaches for Poor Declination Measurements}

From figure \ref{fig:airmass}, we can find that right ascensions are also affected by the stable component of the earth's atmosphere.  In order to evaluate this, we carried out the fitting of the annual parallax with the following six methods:
\begin{enumerate}
\item
	Using five features (feature numbers 2, 4, 5, 6 and 7).  Using all epochs, also including the data points which are far from the declination trends (1st, 5th and 9th epochs for feature number 2;  8th epoch for feature number 5;  5th and 9th epochs for feature number 7) except the 1st and 2nd epochs of feature number 5 which have a long time gap against the later epochs.  Using the data of right ascension only.
	\label{allfive}
\item
	Using three features which have trends in the declination measurements (feature numbers 2, 5 and 7).  Using all epochs, also including the data points which are far from the declination trends (1st, 5th and 9th epochs for feature number 2;  8th epoch for feature number 5;  5th and 9th epochs for feature number 7) except the 1st and 2nd epochs of feature number 5 which have a long time gap against the later epochs.  Using the data of right ascension only.
	\label{allthree}
\item
	Using three features which have trends in the declination measurements (feature numbers 2, 5 and 7).  Excluding the epochs at which the declination data are far from the trends (1st, 5th and 9th epochs for feature number 2;  8th epoch for feature number 5;  5th and 9th epochs for feature number 7) and the 1st and 2nd epochs of feature number 5 which have a long time gap against the later epochs.  Using the data of right ascension only.  This is the method we adopt at \S\ref{section:fitting}.
	\label{smallfit}
\item
	Same as the method No.\ \ref{smallfit}, but using the data of both right ascention and declination.
	\label{small-radec}
\item
	Using three features which have trends in the declination measurements (feature numbers 2, 5 and 7).  Using all epochs, also including the data points which are far from the declination trends (1st, 5th and 9th epochs for feature number 2;  8th epoch for feature number 5;  5th and 9th epochs for feature number 7) except the 1st and 2nd epochs of feature number 5 which have a long time gap against the later epochs.  For the epochs at which the declination data are far from the trends, we tried to recover the declination by setting the CLEAN boxes at the likely sidelobes from the trends of declination data.  In two epochs, 5th and 9th epochs for feature number 7, we failed this recovery, so we omitted them from the fitting.  Using the data of right ascension only.
	\label{forceCLEAN}
\item
	Same as the method No.\ \ref{forceCLEAN}, but using the data of both right ascention and declination.
	\label{force-radec}
\end{enumerate}

Table \ref{table:various-fit} summarizes the result of these fitting.
\begin{table*}
	\caption{Result of various fittings for the annual parallax}
	\label{table:various-fit}
	\begin{center}
		\begin{tabular}{cllccrrr} \\ \hline
			No. & Method & RA/dec & $\varpi$\footnotemark[$*$] & $\sigma$\footnotemark[$\dagger$] & $N$\footnotemark[$\ddagger$] & $n$\footnotemark[$\S$] & $\nu$\footnotemark[$\|$] \\
			& & & [mas] & [mas] & & & \\ \hline
			\ref{allfive} & five features, all epochs & RA only & $0.714 \pm 0.060$ & 0.212 & 29 & 11 & 18 \\
			\ref{allthree} & three features, all epochs & RA only & $0.624 \pm 0.060$ & 0.182 & 19 & 7 & 12 \\
			\ref{smallfit} & three features, epochs on dec trend & RA only & $0.643 \pm 0.049$ & 0.101 & 13 & 7 & 6 \\
			\ref{small-radec} & three features, epochs on dec trend & RA and dec & $0.553 \pm 0.083$ & 0.181 & 26 & 13 & 13 \\
			\ref{forceCLEAN} & three features, all epochs, forced CLEAN boxes & RA only & $0.726 \pm 0.066$ & 0.177 & 17 & 7 & 10\\
			\ref{force-radec} & three features, all epochs, forced CLEAN boxes & RA and dec & $0.625 \pm 0.095$ & 0.269 & 34 & 13 & 21 \\ \hline
			\multicolumn{8}{@{}l@{}}{\hbox to 0pt{\parbox{180mm}{\footnotesize
				\footnotemark[$*$] Annual parallax
				
				\footnotemark[$\dagger$] Uncertainty in one epoch
				
				\footnotemark[$\ddagger$] Total number of data points
				
				\footnotemark[$\S$] Total number of calculated parameters
				
				\footnotemark[$\|$] Degree of freedom in the fitting
			}\hss}}
		\end{tabular}
	\end{center}
\end{table*}
Basically, all fitting results of annual parallax coincide in the error range.  Because the method No.\ \ref{smallfit} has smaller error than the methods No.\ \ref{allfive}, \ref{allthree} and \ref{forceCLEAN}, right-ascension data also have large errors when their declination data have large errors.  Comparing the erros of methods No.\ \ref{smallfit} and \ref{forceCLEAN}, we find that we can not recover these large-declination-error data by setting the CLEAN boxes at the likely sidelobes from the trends on declination data.  Comparing method No.\ \ref{smallfit} and \ref{small-radec}, or method No.\ \ref{forceCLEAN} and \ref{force-radec} show that the declination data have larger errors than the right ascension data even when we try to recover declinations with the settings of CLEAN boxes.  In our observations, it might be better not to use the declination data for the fitting of annual parallax.  We adopted the fitting method No.\ \ref{smallfit} because it is the best fit and does not include bad data.

\section{Discussion}

\subsection{Modification of Physical Parameters of Star Forming Regions}

The distance to the target source MSXDC G034.43$+$00.24 becomes less than half of the kinematic distance of 3.7~kpc.  This changes the parameters of the star forming regions in this infrared dark cloud.  We discuss some examples of them.

\citet{bib:Sanhueza+2010} estimates the virial mass of each core from the following equation, assuming the spherical shape of each core and ignoring the magnetic field and external forces:
\begin{equation}
	\frac{M_{\rm vir}}{\MO} = B \left( \frac{R}{\rm pc} \right) \left( \frac{\Delta v}{\rm km~s^{-1}} \right)^2,
\end{equation}
where $\Delta v$ is the mean line width, $R$ is the radius of the cloud, $B$ is the constant depending on the density profile of the cloud.  They calculated with $B = 210$, corresponding to the uniform density.  Virial mass $M_{\rm vir}$ is proportional to the distance, because $M_{\rm vir}$ is proportional to the radius.

They also calculated the LTE (local thermodynamic equilibrium) masses from the observations of the \atom{C}{}{13}O($3 \rightarrow 2$) and C\atom{O}{}{18}($3 \rightarrow 2$) lines.  Assuming that C\atom{O}{}{18}($3 \rightarrow 2$) is emitted under the LTE condition and this emission is optically thin, LTE mass $M_{\rm LTE}$ is calculated from the following formula:
\begin{eqnarray}
	\frac{M_{\rm LTE}}{\MO}
	& = & 0.565 \left( \frac{\mu_m}{2.72 m_{\rm H}} \right) \left( \frac{[{\rm H}_2/{\rm C}\atom{O}{}{18}]}{3.8 \times 10^6} \right) \left( \frac{D}{\rm kpc} \right)^2 \nonumber \\
	& & \frac{(T_{\rm ex} + 0.88) \exp(15.81 / T_{\rm ex})}{1 - \exp(-15.81 / T_{\rm ex})} \iint \tau_{18} dv d\Omega,
\end{eqnarray}
where $\mu_m$ is the mean molecular weight per ${\rm H}_2$ molecule, $m_{\rm H}$ is the mass of the hydrogen atom, $[{\rm H}_2/{\rm C}\atom{O}{}{18}]$ is the abundance ratio of C\atom{O}{}{18} relative to ${\rm H}_2$, $D$ is the distance, $T_{\rm ex}$ is the excitation temperature, which is employed to be $T_{\rm ex} = 30$~K, $\tau_{18}$ is the optical depth of the C\atom{O}{}{18}($3 \rightarrow 2$) line, $v$ is the radial velocity in the unit of ${\rm km~s}^{-1}$, $\Omega$ is the solid angle in the unit of ${\rm arcmin}^2$.  They derived $\tau_{18}$ from the ratio of the observed brightness temperatures of C\atom{O}{}{18}($3 \rightarrow 2$) to \atom{C}{}{13}O($3 \rightarrow 2$), assuming $[ \atom{C}{}{13}\atom{O}{}{} / \atom{C}{}{}\atom{O}{}{18} ] = 7.6$.  LTE mass is proportional to the square of the distance.

\citet{bib:Rathborne+2006} estimated the dust mass from the 1.2~mm continuum observations with the following equation:
\begin{equation}
	M_{\rm dust} = \frac{F_{\nu} D^2}{\kappa_{\nu} B_{\nu}(T_{\rm dust})},
\end{equation}
where $M_{\rm dust}$ is dust mass, $F_{\nu}$ is the observed flux density, $D$ is the distance, $\kappa_{\nu}$ is the dust opacity per unit mass, $B_{\nu}(T_{\rm dust})$ is the Planck function at the dust temperature $T_{\rm dust}$.  They adopted $\kappa_{\rm 1.2mm} = 1.0~{\rm cm}^2{\rm g}^{-1}$, gas to dust mass ratio of 100, and the dust temperature of $T_{\rm dust} = 15$~K.  Dust mass is also proportional to the square of the distance.

Table \ref{table:param} shows the masses calculated from the previous distance $D = 3.7$~kpc and our distance $D = 1.56$~kpc.  The resultant masses become one order small, and the most massive core is no more than thousand $\MO$.
\begin{table*}
	\caption{Physical parameters of the star forming region MSXDC G034.43$+$00.24}
	\label{table:param}
	\begin{center}
		\begin{tabular}{lrrrrrrrrc} \hline
			& \multicolumn{4}{c}{Values at the Kinematic} & \multicolumn{4}{c}{Values at Our Distance} & \\
			& \multicolumn{4}{c}{Distance of $D = 3.7$~kpc} & \multicolumn{4}{c}{of $D = 1.56$~kpc} & Reference \\
			Millimeter Core & MM1 & MM2 & MM3 & MM4
				& MM1 & MM2 & MM3 & MM4 & \\ \hline
			Virial Mass & 1130 & 1510 & 1370 & $\cdots$
				& 476 & 637 & 578 & $\cdots$ & (1) \\
			LTE Mass & 330 & 1460 & $\cdots$ & $\cdots$
				& 59 & 260 & $\cdots$ & $\cdots$ & (1) \\
			Dust Mass & 1187 & 1284 & 301 & 253
				& 211 & 228 & 54 & 45 & (2) \\ \hline
			$L_{\rm bol}$ & 32000 & $\cdots$ & 9000 & 12000 
				& 5700 & $\cdots$ & 1600 & 2100 & (3) \\
			Spectral Type & O9.5 & $\cdots$ & B0.5 & B0.5
				& B1 & $\cdots$ & B3 & B2 & (3) \\ \hline
			\multicolumn{6}{@{}l@{}}{\hbox to 0pt{\parbox{180mm}{\footnotesize
				The unit of masses and luminosities are $\MO$ and $\LO$, respectively. \par\noindent
				Reference : (1) \citet{bib:Sanhueza+2010}, (2) \citet{bib:Rathborne+2006}, (3) \citet{bib:Rathborne}
			}\hss}}
		\end{tabular}
	\end{center}
\end{table*}

\citet{bib:Rathborne} derived the bolometric luminosities from the standard graybody fitting of the spectral energy distribution (SED) between the mid infrared and millimeter wave.  They assumed that the cores are isothermal and that their radii are \timeform{15"}.  The parameters derived from the fitting are the emissivity index $\beta$, the optical depth at 250~\micron\ $\tau_{250\micron}$, the dust temperature $T_{\rm dust}$, and the bolometric luminosity $L_{\rm bol}$.  Furthermore, the luminosity is constant for the high-mass protostar during the evolution from the protostar to the main-sequence, so the bolometric luminosities tell us the corresponding spectral types of protostars.  These values are also shown in table \ref{table:param}.  They are still early B type star with the mass $\sim 10 \MO$.

\subsection{The Position and Motion on the Galactic Plane}

\begin{figure*}
	\begin{center}
		\FigureFile(140mm, 100mm){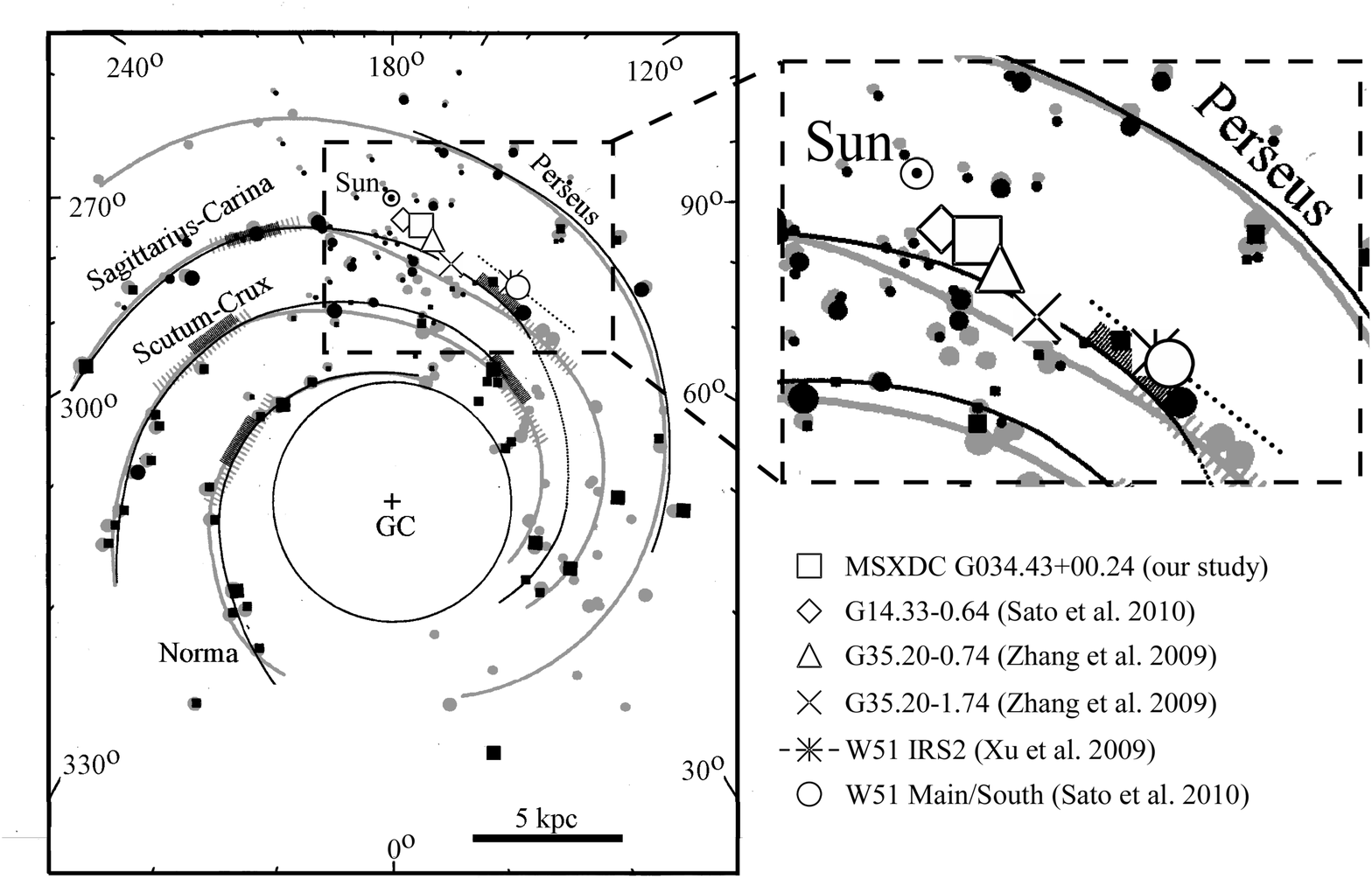}
	\end{center}
	\caption{Positions on the Galactic plane of some sources VLBI parallaxes are measured, overlaid the the model of the Galaxy by \citet{bib:GeorgelinGeorgelin1976} (black) and \citet{bib:TaylorCordes1993} (gray).  ``GC'' is the position of Galactic center.  The open square indicates the position of MSXDC G034.43$+$00.24 based on our annual parallax measurements.  The position of G14.33$-$0.64 (open diamond) is calculated from the parallax measured by \citet{bib:Sato+2010} from the ${\rm H}_2{\rm O}$ maser observations with VERA.  The position of G35.20$-$0.74 (open triangle), G35.20$-$1.74 (cross) and W51 IRS2 (asterisk) are also based on the parallaxes by \citet{bib:Zhang+2009} and \citet{bib:Xu+2009} with the VLBA 12-GHz methanol maser observations.  The position of W51 Main/South (open circle) is based on the parallax measurements by \citet{bib:Sato-astroph} with ${\rm H}_2{\rm O}$ maser observations with the VLBA.  Uncertainty of distances to W51 IRS2 is shown by the dotted line.  The distance uncertainty for the other sources is too small to be hidden by the symbols.  [This figure is made from the figure in \citet{bib:Sato+2010}.]}
	\label{fig:gal-plane}
\end{figure*}

MSXDC G034.43$+$00.24 is located at the closer part of Sagittarius-Carina arm.  Figure \ref{fig:gal-plane} shows the position of MSXDC G034.43$+$00.24 at the distance of 1.4~kpc derived from our annual parallax.
The background is cited from \citet{bib:Sato+2010}.  The black lines and points are the model of the Galaxy by \citet{bib:GeorgelinGeorgelin1976} and the gray by \citet{bib:TaylorCordes1993}, which is modified for the better fit of the kinematic distances by \citet{bib:Downes+1980}.  \citet{bib:TaylorCordes1993} introduced a ``bump'' structure in Sagittarius arm, but our parallax measurement does not support it.  As \citet{bib:Sato+2010} proposed, the ``bump'' may be caused by the errors of kinematic distances.

It might be possible that MSXDC G034.43$+$00.24 has a large peculiar motion in our Galaxy.  When we assume the flat rotation model, the Galactic rotation $V_0 = 220 {\rm ~km~s}^{-1}$ and the distance to the Galactic center $R_0 = 8.5$~kpc, the radial velocity of this source is calculated to be $v_{\rm LSR} = 19 {\rm ~km~s}^{-1}$.  The observed radial velocity of the C\atom{O}{}{18} line is about $58 {\rm ~km~s}$ \citep{bib:Rathborne}, which is less than the calculated velocity by about $40 {\rm ~km~s}^{-1}$.  This might be the line-of-sight component of the peculiar motion.  Furthermore, comparing with other star-forming regions at almost the same galactic latitude such as G35.20$-$0.74 and G35.20$-$1.74, whose distances are measured from parallaxes by \citet{bib:Zhang+2009} of $2.19^{+0.24}_{-0.20}$~kpc and $3.27^{+0.56}_{-0.42}$~kpc, respectively, MSXDC G034.43$+$00.24 is closer to the Sun.  More annual parallax measurements of Sagittarius-Carina-arm sources with VERA reveal the structure of Sagittarius-Carina arm.

\section{Summary}

We carried out the phase-referencing VLBI observations of an infrared dark cloud MSXDC G034.43$+$00.24 with VERA, and derived its annual parallax of $0.643 \pm 0.049$~mas.  This corresponds to the distance of $1.56^{+0.12}_{-0.11}$~kpc, which is less than the half of the previous kinematic distance of 3.7~kpc.  We showed that we are able to measure the parallaxes of low-declination source at $\delta_{\rm J2000} \sim + \timeform{1D}$ with the fitting of right-ascension data only.  The stable components of the earth's atmosphere cause the large scatter in the declination direction of our result.

Since the distance becomes shorter, the masses of millimeter cores decrease by one order of magnitude.  The mass of the most massive millimeter core changes from the previous estimations of $\sim 1000 \MO$ to hundreds $\MO$.  The spectral types derived from the luminosity are B1--B3, which are still early B types.  The position on Galactic plane is the closer part of Sagittarius-Carina arm.  The radial velocity calculated from the flat rotation model has a large offset from observed radial velocities by around $40~{\rm km~s}^{-1}$, which might be the line-of-sight component of the peculiar motion.

\appendix

\section{Details of Data Reduction}
\label{section:analysis}

\begin{figure*}
	\begin{center}
		\FigureFile(160mm, 190mm){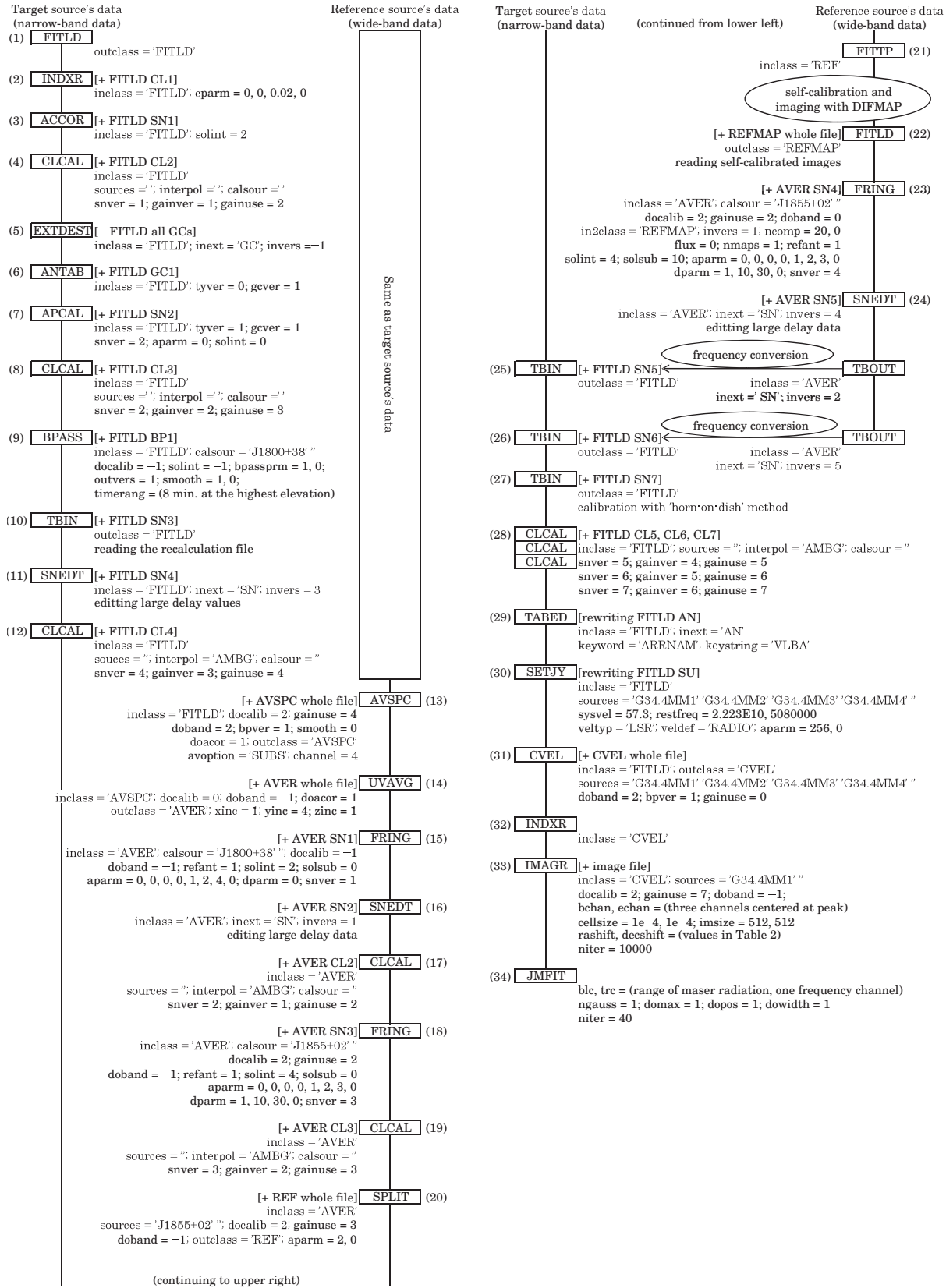}
	\end{center}
	\caption{Details of the data reduction for an observing epoch.  Names in the boxes denote AIPS tasks and verbs.  Some important adverbs, which are the setting parameters for AIPS tasks and verbs, are shown below the task names in the format of ``(adverb name) $=$ (adverb values)''.  Square brackets ([ \ldots ]s) show the outputs of the tasks or verbs. Plus signs mean creating new AIPS tables or files.  Minus signs mean deleting AIPS tables.  AIPS files are denoted with the CLASS names only because NAME parts are used to distinguish observing epochs.}
	\label{fig:analysis}
\end{figure*}

Here, we describe the details of our data reduction.  Figure \ref{fig:analysis} shows the procedure of our data reduction and the setting parameters in the AIPS software.

In steps (1) and (2), we read raw data from correlator (FITS files) to AIPS.  The AIPS task \texttt{INDXR} (Hereafter, the \texttt{CAPITALS OF TYPED FONTS} mean the task or verb (programs) names of AIPS.) was needed for the preparation in the AIPS program.  \texttt{cparm = 0, 0, 0.02, 0} denotes that the time interval of data calibration is 0.02 minutes. [Hereafter, \texttt{small letters of typed fonts} denote the adverbs (setting parameters) of AIPS in the format of `(adverb names) = (setting value)'.]

Steps from (3) to (8) were the amplitude calibration for all sources. \texttt{ACCOR} (3) makes the normalizations of visibilities using auto-correlation data.  The intervals and integrated durations of this solution were set to 2 minutes with \texttt{solint = 2}.  After applying this result with \texttt{CLCAL} (4), we carried out the amplitude calibration with the system temperatures and antenna efficiencies.  We used system temperatures supplied from Mitaka correlator side.  Some antenna efficiencies for different separation angles are supplied from Mitaka correlator, but we used antenna efficiencies at the zero separation angle because there is little change with separation angles\footnote{VERA Status Report, available from \texttt{http://veraserver.mtk.nao.ac.jp/restricted/index-e.html}}.  We deleted all gain-curve (GC) tables (\texttt{inext = GC} and \texttt{invers = -1}) with \texttt{EXTDEST} (5) and read new one from the text file of the zero separation angle with \texttt{ANTAB} (6).  The amplitude calibration itself was done with \texttt{APCAL} (7).  This result was also applied with \texttt{CLCAL} (8).

Bandpass calibration was done with \texttt{BPASS} (9).  The calibration was applied to all sources by using the data of the calibrator source, QSO J1800$+$3848 (\texttt{calsour = 'J1800+38'}) during 8 minutes at the highest elevation (\texttt{solint = -1} and \texttt{timerang = (8 min.\ at the highest elevation)}).  We used raw data of the calibrator source without any calibrations (\texttt{docalib = -1}).  \texttt{bpassprm = 1, 0} means that this calibration was done only for the amplitudes, not for the phases.  Ther is little phase variation within and between the IF(s) because VERA uses the digital filter bank for frequency fileterings.  \texttt{smooth = 1, 0} denotes that we apply Hanning smoothing function, whose width is four channels, before bandpass calibration.

Modification of tracking model in Mitaka FX correlator was done by reading a text file with \texttt{TBIN} (10).  This text file contains the difference of tracking model between Mitaka FX correlator and CALC3/MSOLV, as explained in \S\ref{section:rough-analysis}.  We edited large delay data with \texttt{SNEDT} (11).  This large delays were introduced because of the bug of the program making text files for \texttt{TBIN}.  This calibration was applied with \texttt{CLCAL} (12).

Time and frequency integrations were conducted at steps (13) and (14).  First frequency integration was carried out with \texttt{AVSPC}.  Four frequency channels were integrated to one frequency channel (\texttt{avoption = 'SUBS'} and \texttt{channel = 4}).  This task created new AIPS files, whose visibilities are calibrated with the previous calibrations (\texttt{docalib = 2}, \texttt{gainuse = 4}, \texttt{doband = 2} and \texttt{bpver = 1}).  \texttt{doacor = 1} means that the auto-correlation data are also included in the new AIPS files.  Time integration was conducted with \texttt{UVAVG}.  Data of 4 ses were integrated to one time step (\texttt{xinc = 1}, \texttt{yinc = 4} and \texttt{zinc = 1}).

Calibration of clock parameters was done with the global fringe search [\texttt{FRING}, (15)] of the calibrator source (\texttt{calsour = 'J1800+38' ''}).  The integration time and the time intervals of solutions were 2 minutes (\texttt{solint = 2}, \texttt{solsub = 0}).  The detection threshold was 4 in the signal-to-noise ratio (\texttt{aparm = 0, 0, 0, 0, 1, 2, 4, 0}).  After editing large delay data with \texttt{SNEDT} (16), the results were applied with \texttt{CLCAL} (17), interpolating $2\pi$ ambiguity of phases using rate values (\texttt{interpol = 'AMBG'}).  Then the global fringe search [\texttt{FRING}, (18)] was conducted for the reference source (\texttt{calsour = 'J1855+02' ''}).  We set the delay and window of 10 ns and 30 mHz, respectively (\texttt{dparm = 1, 10, 30, 0}).  This result was also applied with \texttt{CLCAL} (19).  After integrating all frequency channels of each IF with \texttt{SPLIT} (20) (\texttt{aparm = 2, 0}) and outputting to FITS file with \texttt{FITTP} (21), the reference source was imaged with self-calibration using DIFMAP.

For phase referencing, we reran the global fringe search for the reference source.  After loading the image of the reference source with \texttt{FITLD} (22), we carried out the global fringe search with \texttt{FRING} (23).  The image is used for the model of the source structure (\texttt{ncomp = 20, 0}, \texttt{nmaps = 1}).  The integration time and the solution intervals were 4 minutes and 0.4 minutes, respectively (\texttt{solint = 4}, \texttt{solsub = 10}).  Editing large delay data with \texttt{SNEDT} (24), this result and the result of calibration of clock parameters were transferred to the target source data (25, 26).  Here we needed to convert phases because they depend on frequencies in $\phi = 2\pi\nu\tau$, where $\phi$, $\nu$ and $\tau$ are phase, frequency and delay.  After loading the calibration of `horn-on-dish' method (27), these results were applied (28).

Calibration of Doppler effect in VLBI does not include the earth's spin.  AIPS conducts this calibration only when the ARRNAM (array name) keyword in antenna tables is VLBA.  So we changed it from VERA to VLBA with task \texttt{TABED} (29).  The parameters of Doppler calibration were set with \texttt{SETJY} by rewriting source (SU) tables (30).  The frequency channel of $v_{\rm LSR} = 57.3 {\rm ~km~s}^{-1}$ was moved to the channel number 256 (\texttt{sysvel = 57.3}, \texttt{veltyp = 'LSR'}, \texttt{aparm = 256, 0}).  The rest frequency was set to 22.23508 GHz (\texttt{restfreq = 2.223E10, 5080000}).  The calibration itself was conducted with \texttt{CVEL} (31) for the target sources (\texttt{sources = 'G34.4MM1', 'G34.4MM2', 'G34.4MM3', 'G34.4MM4' ''}).  \texttt{INDXR} (32) needed for creating AIPS internal files deleted by \texttt{CVEL}.

Imaging of target source was carried out with \texttt{IMAGR} (33).  The imaged source was MM1 (\texttt{sources = 'G34.4MM1' ''}).  The size of a pixel was $(1 \times 10^{-4} {\rm ~mas}) \times (1 \times 10^{-4} {\rm ~mas})$ (\texttt{cellsize 1e-4, 1e-4}).  The resultant images had $512 \times 512$~pixels (\texttt{imsize = 512, 512}).  The maximum number of clean components were 10000 (\texttt{niter = 10000}).  Measurement of positions and fluxes was done by fitting elliptical Gaussians with \texttt{JMFIT} (34).  The maximum step of iteration in the fitting was 40 (\texttt{niter = 40}).



\begin{thebibliography}{}
\bibitem[Asaki \etal(2007)]{bib:Asaki}
	Asaki,~Y., \etal\ 2007, \pasj, 59, 397
\bibitem[Baba \etal(2009)]{bib:baba}
	Baba,~J., Asaki,~Y., Makino,~J., Miyoshi,~M., Saitoh,~T.~R., \& Wada,~K. 2009, \apj, 706, 471
\bibitem[Bronfman, Nyman, May(1996)]{bib:Bronfman+1996}
	Bronfman,~L., Nyman,~L.-\AA, \& May,~J. 1996, \aaps, 115, 81
\bibitem[Carey \etal(1998)]{bib:Carey+1998}
	Carey,~S.~J., Clark,~F.~O., Egan,~M.~P., Price,~S.~D., Shipman,~R.~F., \& Kuchar,~T.~A. 1998, \apj, 508, 721
\bibitem[Carey \etal(2000)]{bib:Carey+2000}
	Carey,~S.~J., Feldman,~P.~A., Redman,~R.~O., Egan,~M.~P., MacLeod,~J.~M., \& Price,~S.~D. 2000, \apj, 543, L157
\bibitem[Chambers \etal(2009)]{bib:Chambers+2009}
	Chambers,~E.~T., Jackson,~J.~M., Rathborne,~J.~M., \& Simon,~R. 2009, \apjs, 181, 360
\bibitem[Charlot(1990)]{bib:structure2}
	Charlot,~P. 1990, \aj, 99, 1309
\bibitem[Downes \etal(1980)]{bib:Downes+1980}
	Downes,~D., Wilson,~T.~L., Bieging,~J. \& Wink,~J. 1980, \aaps, 40, 379
\bibitem[Egan \etal(1998)]{bib:Egan+1998}
	Egan,~M.~P., Shipman,~R.~F., Price,~S.~D., Carey,~S.~J., Clark,~F.~O., \& Cohen,~M. 1998, \apj, 494, L199
\bibitem[Fa{\'u}ndez \etal(2004)]{bib:Faundez+2004}
	Fa{\'u}ndez,~S., Bronfman,~L., Garay,~G., Chini,~R., Nyman,~L.-\AA., \& May,~J. 2004, \aap, 426, 97
\bibitem[Fomalont \etal(2003)]{bib:Fomalont+2003}
	Fomalont,~E.~B., Petrov,~L., MacMillan,~D.~S., Gordon,~D. \& Ma,~C. 2003, \aj, 126, 2562
\bibitem[Fujishita(1983)]{bib:structure1}
	Fujishita,~M. 1983, Publ. Int. Latitude Obs. Mizusawa, 17, 13
\bibitem[Georgelin, Georgelin(1976)]{bib:GeorgelinGeorgelin1976}
	Georgelin,~Y.~M. \& Georgelin,~Y.~P. 1976, \aap, 49, 57
\bibitem[Garay \etal(2004)]{bib:Garay+2004}
	Garay,~G., Fa{\'u}ndez,~S., Mardones,~D., Bronfman,~L., Chini,~R., \& Nyman,~L.-\AA. 2004, \apj, 610, 313
\bibitem[Green(1985)]{bib:Green1985}
	Green,~R.~M. 1985, Spherical Astronomy (Cambridge : Cambridge University Press)
\bibitem[Hennebelle \etal(2001)]{bib:Hennebelle+2001}
	Hennebelle,~P., P\'{e}rault,~M., Teyssier,~D., \& Ganesh,~S. 2001, \aap, 365, 598
\bibitem[Honma \etal(2008)]{bib:2bcal}
	Honma,~M., \etal\ 2008, \pasj, 60, 935
\bibitem[Honma, Tamura, Reid(2008)]{bib:secz}
	Honma,~M., Tamura,~Y., \& Reid,~M.~J. 2008, \pasj, 60, 951
\bibitem[Iguchi \etal(2005)]{bib:DFU}
	Iguchi,~S., Kurayama,~T., Kawaguchi,~N., \& Kawakami,~K. 2005, \pasj, 57, 259
\bibitem[Jackson \etal(2008)]{bib:Jackson+2008}
	Jackson,~J.~M., Finn,~S.~C., Rathborne,~J.~M., Chambers,~E.~T., \& Simon,~R. 2008, \apj, 680, 349
\bibitem[Jike \etal(2005)]{bib:JikeFXCALC}
	Jike,~T., Fukuzaki,~Y., Shibuya,~K., Doi,~K., Manabe,~S., Jauncey,~D.~L., Nicolson,~G.~D., \& McCulloch,~P.~M. 2005, Polar Geosci., 18, 26
\bibitem[Kamohara \etal(2010)]{bib:Kamohara+2010}
	Kamohara,~R., \etal\ 2010, \aap, 510, A69
\bibitem[Kawaguchi, Sasao, Manabe(2000)]{bib:2BKawaguchi}
	Kawaguchi,~N., Sasao,~T., \& Manabe,~S. 2000, Proc. SPIE, 4015, 544
\bibitem[Kurayama, Sasao, Kobayashi(2005)]{bib:Kurayama2005}
	Kurayama,~T., Sasao,~T., \& Kobayashi,~H. 2005, \apj, 627, L49
\bibitem[Manabe, Yokoyama, Sakai(1991)]{bib:ManabeFXCALC}
	Manabe,~S., Yokoyama,~K., \& Sakai,~S. 1991, IERS Techn. Note, 8, 61
\bibitem[Miralles, Rodr{\'\i}gues, Scalise(1994)]{bib:Miralles+1994}
	Miralles,~M.~P., Rodr{\'\i}gues,~L.~F., \& Scalise,~E. 1994, \apjs, 92, 173
\bibitem[Molinari \etal(1998)]{bib:Molinari+1998}
	Molinari,~S., Brand,~J., Cesaroni,~R., Palla,~F., \& Palumbo,~G.~G.~C. 1998, \aap, 336, 339
\bibitem[Motogi \etal(2011)]{bib:Motogi+}
	Motogi,~K., Sorai,~K., Habe,~A., Honma,~M., Kobayashi,~H., \& Sato,~K. 2011, \pasj, 63, 31
\bibitem[Oh \etal(2010)]{bib:Oh+2010}
	Oh,~C.~S., Kobayashi,~H., Honma,~M., Hirota,~T., Sato,~K., \& Ueno,~Y. 2010, \pasj, 62, 101
\bibitem[P{\'e}rault \etal(1996)]{bib:Perault+1996}
	P{\'e}rault,~M., \etal\ 1996, \aap, 315, L165
\bibitem[Pillai \etal(2006)]{bib:Pillai+2006}
	Pillai,~T., Wyrowski,~F., Carey,~S.~J., \& Menten,~K.~M. 2006, \aap, 450, 569
\bibitem[Rathborne \etal(2005)]{bib:Rathborne}
	Rathborne,~J.~M., Jackson,~J.~M., Chambers,~E.~T., Simon,~R., Shipman,~R., \& Frieswijk,~W. 2005, \apj, 630, L181
\bibitem[Rathborne, Jackson, Simon(2006)]{bib:Rathborne+2006}
	Rathborne,~J.~M., Jackson,~J.~M., \& Simon,~R. 2006, \apj, 641, 389
\bibitem[Rathborne, Simon, Jackson(2007)]{bib:Rathborne+2007}
	Rathborne,~J.~M., Simon,~R., \& Jackson,~J.~M. 2007, \apj, 662, 1082
\bibitem[Rygl \etal(2010)]{bib:Rygl+2010}
	Rygl,~K.~L.~J., Brunthaler,~A., Reid,~M.~J., Menten,~K.~M., van~Langevelde,~H.~J., \& Xu,~Y. 2010, \aap, 511, A2
\bibitem[Sato \etal(2010a)]{bib:Sato+2010}
	Sato,~M., Hirota,~T., Reid,~M.~J., Honma,~M., Kobayashi,~H., Iwadate,~K., Miyaji,~T., \& Shibata,~K.~M. 2010a, \pasj, 62, 287
\bibitem[Sato \etal(2010b)]{bib:Sato-astroph}
	Sato,~M., Reid,~M.~J., Brunthaler,~A., \& Menten,~K.~M. 2010b, \apj, 720, 1055
\bibitem[Sanhueza \etal(2010)]{bib:Sanhueza+2010}
	Sanhueza,~P., Garay,~G., Bronfman, L., Mardones,~D., May,~J., \& Saito,~M. 2010, \apj, 715, 18
\bibitem[Shepherd \etal(2007)]{bib:Shepherd+2007}
	Shepherd,~D.~S., \etal\ 2007, \apj, 669, 464
\bibitem[Simon \etal(2006a)]{bib:Simon+2006a}
	Simon,~R., Jackson,~J.~M., Rathborne,~J.~M., \& Chambers,~E.~T. 2006a, \apj, 639, 227
\bibitem[Simon \etal(2006b)]{bib:Simon+2006b}
	Simon,~R., Rathborne,~J.~M., Shah,~R.~Y., Jackson,~J.~M., \& Chambers,~E.~T. 2006b, ApJ, 653, 1325
\bibitem[Taylor, Cordes(1993)]{bib:TaylorCordes1993}
	Taylor,~J.~H. \& Cordes,~J.~M. 1993, \apj, 411, 674
\bibitem[Tayssier, Hennebelle, P{\'e}rault(2002)]{bib:Tayssier+2002}
	Tayssier,~D., Hennebelle,~P., \& P{\'e}rault,~M. 2002, \aap, 382, 624
\bibitem[Thompson, Moran, Swenson(2001)]{bib:Thompson}
	Thompson,~A.~R., Moran,~J.~M., \& Swenson,~G.~W.,~Jr. 2001, Interferometry and Synthesis in Radio Astronomy, 2nd edition (New York: John Wiley \& Sons)
\bibitem[Wang \etal(2006)]{bib:VLA}
	Wang,~Y., Zhang,~Q., Rathborne,~J.~M., Jackson,~J., \& Wu,~Y.\ 2006, \apj, 651, L125
\bibitem[Xu \etal(2009)]{bib:Xu+2009}
	Xu,~Y., Reid,~M.~J., Menten,~K.~M., Brunthaler,~A., Zheng,~X.~W. \& Moscadelli, L. 2009, \apj, 693, 413
\bibitem[Zhang \etal(2009)]{bib:Zhang+2009}
	Zhang,~B., Zheng,~X.~W., Reid,~M.~J., Menten,~K.~M., Xu,~Y., Moscadelli,~L. \& Brunthaler,~A. 2009, \apj, 693, 419
\end{thebibliography}
\end{document}